\documentclass[runningheads]{llncs}
\usepackage[dvips]{graphicx}

\usepackage{amssymb}
\def\Int{\mathrm{int}}

\def\clique{\mathrm{Clique}}
\def\Int{\mathrm{Int}}
\def\Gdag{\mathrm{\overrightarrow{G}}}
\def\GOdag{\mathrm{\overrightarrow{GO}}}

\newcommand{\ind}{\mbox{\rm ind}}

\newcommand{\Floor}[1]{\mathrm{\lfloor #1\rfloor}}

\begin{document}
\mainmatter

\title{Indexing of Tables Referencing Complex Structures}

\titlerunning{Indexing Complex Structures.}

\author{Agust S.~Egilsson\inst{1}
\and Hakon Gudbjartsson\inst{2}}

\institute{
Department of Mathematics,
University of California,
Berkeley, CA\\
\email{Egilsson@Math.Berkeley.EDU}
\and
deCODE Genetics Inc.,
Sturlugata  8,
IS-101 Reykjavik,
Iceland \\
\email{Hakon@deCODE.is}}

\authorrunning{Egilsson et Gudbjarsson.}

\maketitle

\begin{abstract}
We introduce indexing of tables referencing complex structures such as digraphs and spatial objects, appearing in genetics and other data intensive analysis. The indexing is achieved by extracting dimension schemas from the referenced structures. The schemas and their dimensionality are determined by proper coloring algorithms and the duality between all such schemas and all such possible proper colorings is established. This duality, in turn, provides us with an extensive library of solutions when addressing indexing questions. It is illustrated how to use the schemas, in connection with additional relational database technologies, to optimize queries conditioned on the structural information being referenced. Comparisons using bitmap indexing in the Oracle 9.2i database, on the one hand, and multidimensional clustering in DB2 8.1.2, on the other hand, are used to illustrate the applicability of the indexing to different technology settings. Finally, we illustrate how the indexing can be used to extract low dimensional schemas from a binary interval tree in order to resolve efficiently interval and stabbing queries.

\vspace{2 mm}


\end{abstract}

\begin{multicols}{2}

\section{Introduction}
\label{sec:intro}

Researchers at the Whitehead Institute/MIT Center for Genome Research, use flat file formats when working on whole genome assembly, according to a conversation with David Jaffe the group leader. Is this the rule or an exception? It still appears that flat file formats are currently used in many, if not most, computationally intensive biology related research. To us this represents a challenge: To come up with relational database technologies which allow efficient analysis of complex structures, e.g., from biology, using the relational database environment.

The approach that we are led to in this paper involves generalizing existing ROLAP (Relational OLAP) methods. The generalization is achieved by applying proper graph colorings in order to extract dimension schemas from complex structures. A rather cryptic formula, see the corollary in section \ref{sec:low}, dictates that this generalization, when applied to many acyclic digraphs, will return schemas of low dimensionality. In turn, this low dimensionality suggests that the database - and, in particular, the business intelligence industry is well placed to provide new tools incorporating the generalized methods.

Indexing of tables referencing graphs and other complex structures is studied in this paper. For digraphs the setting is as follows. Start with a table in a relational database. Assume that one of the columns in the table contains nodes from a directed graph. We consider the task of indexing or organizing the table in such a way that queries based on the information provided by the edges of the digraph are optimally evaluated. 

The indexing introduced differs from many other approaches in that it is not complete. In our case this means that only a data entry schema is materialized, as a table, and then additional mechanisms are borrowed from the relational database system. How exactly the data entry schemas, called clique schemas, are used to optimize queries varies, it is determined by the database optimizer. To complete the indexing the additional relational mechanisms used here, for illustration purposes, are static bitmap (join) indexes and multidimensional clustering. This leads us to comparing using clique schemas in connection with multidimensional clustering in DB2, see \cite{IBM-Cluster}, and using the schemas in connection with static bitmap join indexing in Oracle, see \cite{Oracle-Star}. The comparison reveals that the clique indexing schema can be used efficiently in connection with various relational data access methods. 

It is worth pointing out that in the paper ``An Array-Based Algorithm for Simultaneous Multidimensional Aggregates" by Zhao, Deshpande and Naughton (\cite{Zhao}), clustered arrays and non-clustered tables are compared with respect to the CUBE operator. Our setup is somewhat similar but our results do not necessarily favor using multidimensional structures as was observed for the CUBE operator in \cite{Zhao}. In fact, the Oracle setup, combining clique schemas and bitmap join indexing, turns out to be much more efficient when analyzing data residing in memory than using multidimensional clustering in DB2.

Additional examples are also introduced. This includes clique indexing schemas for various set collections other than digraphs, such as structures derived from sets of intervals or the binary interval tree. In the latter, the clique schemas may be used to answer efficiently stabbing and interval queries. The examples show that the technology can be used to address a large class of problems involving complex structures.

\subsection{Chronology}
Section \ref{sec:low} contains a discussion about dimensionality of graph structures. In section \ref{sec:complex} we derive the clique indexing schema using proper graph coloring. The schema is then put in a context in which it provides an optimal solution to determining data entries for indexing tables referencing sets determined by a set-valued function. The indexing is tested in section \ref{sec:comparisons} in connection with other relational database techniques. Additional applications are provided for illustration in section \ref{sec:applications}. In each of the additional applications we use the clique schema to implement spatial indexing, first we use a set-valued function which depends on the underlying spatial data and then we model a fixed interval tree instead.

\section{Related Work}
\label{sec:related}
Indexing techniques for relational databases is a large field of study. An overview of indexing of semistructured data, e.g., acyclic digraphs and XML data, may be found in \cite{Ab-Bu-Su}. A particular indexing technique for digraphs based on encoding paths as strings is introduced in \cite{Gisli}. In \cite{Framework} the authors, of that paper, describe an analysis framework for tree-structured balanced access methods for evaluating performance, based on ``performance loss metrics" and assumed optimal tree structures. Stabbing and interval queries may be resolved using various trees and structures and recently the relational interval tree has been introduces, see \cite{RI} for the RI-tree and further references.

Several relevant graph coloring results, examples and observations regarding down-coloring of acyclic digraphs are obtained in \cite{GeirAgust1} and \cite{GeirAgust2}. In particular, estimates are obtained for the down-chromatic numbers of acyclic digraphs in \cite{GeirAgust2} and performance factors established for the down-coloring of acyclic digraphs. An accessible introduction to graph coloring is found in \cite{DWest}.

\section{Dimensionality}
\label{sec:low}

Today's decision support, also know as business intelligence, industry thrives on promoting technology made possible, to a large degree, by the low dimensionality of digraphs describing business processes. For example, a typical location hierarchy has 4 levels i.e., store, city, state and country. The time dimension can likewise, usually, be colored with few colors and so on. For science related applications the structures are more complicated, but we still may want to be able to take advantage of the efficiency (the online part in OLAP or online-analytical-processing) provided by star-schema etc. like setups, made possible by low chromatic numbers. Dimensionality in OLAP setups generalizes to down-chromatic numbers, defined in the papers \cite{GeirAgust1} and \cite{GeirAgust2}. We will explain this further, but first we review an important, although cryptic, formula from \cite{GeirAgust2}. It is important because it bounds the dimensionality of indexing structures required for acyclic digraphs. Just as in today's business intelligence systems, low dimensionality allows one to materialize the indexing (clique) schemas as dimension tables. 
\medskip

\noindent In \cite{GeirAgust2} the following is proved about the down-chromatic number of an acyclic digraph $\Gdag$.

\begin{corollary}[from \cite{GeirAgust2}]
If $\Gdag$ is an acyclic digraph, then its down-chromatic number
satisfies the following:
\begin{enumerate}
\item If $\ind(H) = 1$ or $D(\Gdag) = 2$, then \\$\chi_d(\Gdag) = D(\Gdag)$.
\item If $\ind(H) > 1$ and $D(\Gdag)> 2$, then \\$\chi_d(\Gdag) \leq \ind(H)(D(\Gdag) - 2) + 1$.
\end{enumerate}
Where $H = H_{\small{\Gdag}}$ is the down-hypergraph of $\Gdag$. Moreover, the mentioned upper bounds in both cases
are sufficient for greedy down-coloring of $\Gdag$.
\end{corollary}

\noindent
Before explaining the relevance of the corollary further we review the terminology used:
\begin{itemize}
\item The down-chromatic number $\chi_d(\Gdag)$ is the least number of ``colors" that can be applied to the graph nodes so that any two nodes which have a common ancestor always receive different colors. 
\item The down-hypergraph $H$ is obtained from $G$ by taking as its edges the {\it maximal} descendants-and-self sets $D[u]$. More explicitly, $D[u]$ contains $u$ and all nodes that can be traversed starting from $u$ in the digraph. Also, hypergraphs are allowed to have edges formed by connecting multiple nodes instead of just 1 or 2 nodes per edge for graphs.
\item The number $\ind(H)$ is called the degeneracy or the inductiveness of the down-hypergraph. It is defined as
\[
\begin{array}{rcl}
\ind(H) &=& \max_{S\subseteq V(H)}\left\{ \delta(H[S])\right\}\\
 &=& \max_{S\subseteq V(H)}\left\{\min_{u\in S}\left\{d_{H[S]}(u)\right\}\right\}
\end{array}
\]
where $V(H)$ is the set of graph nodes (from $\Gdag$) and $H[S]$ is the subhypergraph of $H$ induced by $S$. The subhypergraph contains nodes from $S$ and its edges are obtained as the (distinct) restrictions of maximal descendants-and-self sets $D[u]$ to $S$ and additionally by requiring each edge, in $H[S]$, to be formed by at least two nodes. Finally, to complete the description of $\ind(H)$ we point out that the minimum degree of $H[S]$, denoted by $\delta(H[S])$, represents the absolute {\it minimum} count of edges from $H[S]$ that contain the same element from $S$. 
\item The number $D(\Gdag)$ is the maximum number of nodes in a descendants-and-self set. 
\end{itemize}

The relevance of the corollary here is that the down-chromatic number $\chi_d(\Gdag)$ estimated in the corollary represents the number of dimensions needed to ``index" a table referencing the digraph $\Gdag$. The indexing is used to resolve queries conditioned on unions, intersections and compliments of ancestors-and-self sets derived from the digraph $\Gdag$. Eventually, we will extend our approach to any set-valued mapping. We refer the reader to \cite{GeirAgust2} for a more detailed discussion of these numbers. A high level discussion follows.

When considering the above formulas it is sometimes useful to reverse the arrows in an acyclic digraph so that the nodes converge at a ``root" element, if such an element exists. Then, if the arrows are reversed, $D(\Gdag)$ is the number of nodes in the largest ancestors-and-self set in the original digraph and the down-coloring produces an indexing schema allowing us to optimize queries referencing descendants-and-self sets. In this case, the schemas may therefore be considered to be generalizations of dimension tables found in star-schema setups commonly used in relational databases.

Most importantly, for many applications, including (reversed) tree digraphs, the degeneracy of the down-hypergraph is one, i.e., $\ind(H) = 1$. Therefore, in these cases, only the first part of the corollary needs to be evaluated and the dimensionality or down-chromatic number, i.e., $\chi_d(\Gdag)$, is readily determined, in the simplest way one can hope for, as $D(\Gdag)$. Sometimes, such as for graphs similar to (reversed) genealogy digraphs, $D(\Gdag)$ may grow fast, but eventually lack of information and inbreeding will bound the largest ancestor set. On the other hand, polygamy can result in higher degeneracy numbers, requiring us to use the estimate in part two of the corollary. We finish this section with an example of a small ``artificial" digraph $\Gdag$ with $\ind(H) > 1$ and $D(\Gdag) > 2$. The acyclic digraph $\Gdag$ below has $\ind(H) = 3$, $D(\Gdag)=3$ and $\chi_d(\Gdag) = 4$.

$$\begin{array}{cccccccccccccccc}
      &&&         &   &          & 12     &          &   &\\
      &&&         &   & \swarrow &        & \searrow &   &\\
      &&&         & 1 &          && & 2 &\\
      &&\nearrow& & & \nwarrow &  & \nearrow &  & \nwarrow & \\
      14&&&         &   &&
      \begin{array}{cc}
      13&24\\
      24&13\\
      \end{array}
      && &  & 23 \\
      &&\searrow&   &   & \swarrow & & \searrow &   & \swarrow & \\
      &&&         & 4 &          &       &   & 3 &\\
      &&&         &   & \nwarrow &        & \nearrow &   &\\
      &&&         &   &          & 34     &          &   &&\\
\end{array}$$

\noindent
The digraph is defined so that each node ``AB" is the source node for edges with target nodes ``A" and ``B". It can be down-colored with 4 colors. Other similar examples can be found in \cite{GeirAgust2}.

In the following sections we explain further how to extract schemas, similar to dimension tables, from complex structures, such as digraphs or spatial maps. This will enable us to index tables referencing the structures and take advantage of access plans developed for data warehousing.

\section{Tables Referencing Complex Structures}
\label{sec:complex}

In this section we create a simple theoretical framework for studying the indexing. In particular, the duality between proper colorings and indexing of tables referencing complex structures is established. This duality allows us to take advantage of results known about graph coloring when creating indexing schemas and when maintaining the schemas during structure updates. Proper graph coloring has been studied extensively and is used to address numerous applied problems and therefore this connection provides us with a library of solutions when addressing indexing questions. In the previous section, we already showed how a particular coloring result is translated into information about indexing schemas, providing the reader with a motivating example.

\subsection{Schema Representations}

Data entries in an index file are values that point to records in a data file, in our case to rows in a table referencing a domain $V$. It is assumed that each row contains a node, in a fixed column, from the domain $V$. In order to index the table, one needs to develop a data entry schema compatible with the queries that the index will resolve and a translation mechanism that translates the queries into operations carried out with the aid of the data entries, or more formally: Let the data entries be denoted by $e_1*,\ldots, e_n*$, each data entry points to rows in the table and these rows do in turn determine a set of nodes in $V$ which we denote by $F(e_i*)$, the function $F$ is referred to as being set-valued. The data entry schema may be realized as a collection of functions $\{f\}$ defined on subsets of $V$ and with range in the data entries $\{e_1*,\ldots, e_n*\}$ and such that for a given data entry $e*$ there should be at least one function $f$,
in the schema, satisfying $F(e*) = f^{-1}(e*)$. That is, $f$ assigns the data entry $e*$ only to search key nodes from the set $F(e*)$. Therefore, a complete schema is realized as a collection $f_1, f_2, \ldots, f_k$ of 
functions so that for each data entry $e*$ there exists at least one integer $c(e*)$ with
\[
F(e*) = f_{c(e*)}^{-1}(e*).
\]
For two distinct data entries $e*$ and $e'*$ we have that if the intersection $F(e*)\cap F(e'*)$ is nonempty, say contains a node $w$, then $f_{c(e*)}(w) = e*$ and
$f_{c(e'*)}(w) = e'*$, so since $e* \neq e'*$, we must have
$c(e*) \neq c(e'*)$. It follows that the map $e* \mapsto c(e*)$ is a proper coloring of the intersection graph, $\Int(F)$, of the set-valued map $F$. To clarify this, we note that the intersection graph has nodes $$V(\Int(F)) = \{e_1*,\ldots,e_n*\}$$ and edges $$\begin{array}{rcl}
E(\Int(F)) & = &  \{e_i*\leftrightarrow e_j*~:~ e_i*\neq e_j* \\ && ~\mbox{~and~} F(e_i*) \cap F(e_j*) \neq \emptyset\}.
\end{array}$$ 
Consequently, the number of functions, $k$, needed to realize the data entry schema is greater than or equal to the chromatic number, $\chi(\Int(F))$, of the intersection graph. Keeping the number of functions close to this limit is important, since there are considerable performance and bounding costs associated with each additional function, as will become apparent in the next sections.

Before continuing, it is worth pointing out the duality between proper colorings of the $\Int(F)$ graph and possible data entry realizations: Given a proper coloring $c:V(\Int(F)) \rightarrow
\{1,\ldots,k\}$ of the intersection graph, one constructs a complete schema for assigning data entries by defining functions 
$f_1, \ldots,f_k$ as 
\begin{equation}
\label{eq:f}
f_i(u) = e*, \mbox{ if } u \in F(e*) \mbox{ and } c(e*) = i.
\end{equation}
The definition of $\Int(F)$ ensures that the functions are
well defined. The duality between all the proper colorings and all the possible indexing schemas has therefore been established.

\subsection{Generalized Dimension Tables}
\label{sec:generalized}

We sometimes create the data entry schema without a referencing table. This is possible if the structure of the (maximal) set-valued map $F$ is known beforehand, as is often the case when tables reference fixed digraphs. The schema $f_1, \ldots, f_k$ may be materialized in many different ways. Now we illustrate how to use data entry schemas to build tables, similar, in function, to dimension tables used in data warehousing. These tables are called clique tables or schemas.

As an illustration, consider a table referencing a simple digraph $\Gdag$ and the task (again - see section \ref{sec:low} above), of indexing the table in such a way that queries conditioned on one or more of the descendants-and-self sets of the digraph may be efficiently evaluated. In this case it is straightforward to pick as data entries the digraph nodes $V(\Gdag)$ and have each data entry $e*$ point to all rows in the table that reference $e*$ or one of its descendants, i.e., $D[e*]$. The practicality of this approach depends largely on the chromatic number of the intersection graph, $\Int(D[\cdot])$, and the efficiency of the coloring algorithm select to properly color the intersection graph.

For digraphs that induce relatively small, e.g., less than 1000, chromatic numbers of $\Int(D[\cdot])$, we devise a structure so that the set operations $(\cup, ~\cap, ~\setminus)$ can be optimally executed on elements from the set collection $\{D[u]~:~ u\in V(\Gdag)\}$. This is done by having the data entry schema functions, see formula (\ref{eq:f}) above, physically share the domain $V(\Gdag)$ in the database, such as by materializing the relation
\[
\{(u,f_1(u),\ldots,f_k(u)): u\in V(\Gdag)\}
\]
in the database system, in addition to the coloring map $c$. Following standard convention, this requires that
$f_j(u) =$ ``NULL" if $u$ is not in the domain of $f_j$. We will refer to this relation in the below as $\clique(D[\cdot])$, it has ordered column headings: ``node", ``c1", \ldots, ``ck". The data entry schema is referred to as the clique indexing schema. In general, we let
$$\begin{array}{rcl}
\clique(F) &=& [\mbox{``node",~``c1",~...,~``ck"}]\\
&=& \{(u,f_1(u),\ldots,f_k(u)): u\in V\},
\end{array}
$$ representing a proper coloring of the intersection graph $\Int(F)$ and, at the same time, the schema $f_1, \ldots, f_k$ as determined by formula (\ref{eq:f}).

In some cases, such as when working with genealogy records that stretch many generations back, the chromatic number of the intersection graph may become too large for the clique schema to be formed directly. In these cases, one tries to invent a more practical set-valued function, and use it to build the clique schema, instead of using the descendants-and-self mapping directly. This will become apparent in the following sections in connection with the discussion of the interval graph.

\section{Multidimensional Clustering or Bitmap Join Indexing?}
\label{sec:comparisons}

In this section we compare different relational database technologies required to complete the indexing. We apply the indexing to a test setup involving a table referencing a digraph originating from genetics or drug related research. The test setup involves queries resolved using the $\clique(D[\cdot])$ indexing schema, created for the Gene Ontology (denoted by $\GOdag$) digraph. The timing and bounding results are detailed in appendix \ref{sec:appA}.

\subsection{Comparison} 

The clique indexing schema describes mappings of data entries, i.e., pointers, to rows in tables referencing the $\GOdag$ diagraph. Additional relational mechanisms are introduced to materialize the actual mapping of the data entries to RID-s (unique row locations) in the referencing tables. We compare the performance of two different approaches:

\begin{enumerate}
\item Use bitmap join indexing in the Oracle 9.2i database system to map data entries in the clique indexing schema to RID-s in a referencing table. Static bitmap indexes are essentially pointers, represented as compressed bit arrays, to sets of ordered rows in a table. Bitmap indexes therefore behave nicely with respect to logical operations and table lookups. These and a few other data warehousing indexing techniques are discussed in \cite{Variant}.

\item Use multidimensional clustering in DB2 version 8.1.2 to index (using block indexes) and cluster a referencing table so that all rows referencing the same descendants-and-self set are always clustered together. Currently this is only supported, in DB2, for 16 color columns simultaneously so this was applied to a subset of the nodes only. We compare clustering by 1, 2, 4, 8 and 16 color columns in the DB2 setup. Multidimensional clustering organizes tables according to preset column dimensions into chunks of data, physically located together on one or more pages, i.e., blocks, and according to a uniform treatment for all the organizing dimensions. Clustering of arrays is described in \cite{Sarawagi}.  
\end{enumerate}

Not too surprisingly, it turns out that multidimensional clustering performs well when all the referencing data is retrieved from a disk storage, whereas bitmap join indexing outperforms by a huge ratio when a large part of the data is located in memory.

\subsection{Conclusions}

Figure \ref{fig1} in appendix \ref{sec:appA}, demonstrates that both technologies, multidimensional clustering and bitmap join indexing, are useful in the context of indexing tables referencing complex structures when the indexing is facilitated by the clique data entry schema.

Explicitly, the timing of the queries shows the benefits of using multidimensional clustering in DB2 when the chromatic number of the associated intersection graph of the referenced structure is very small, i.e., 16 or less. Having to horizontally append the fact table in DB2 with the color columns is a cumbersome step though, especially if the referenced structure changes with time. Another observation is that the size of the fact table, in DB2, will grow considerably if many color columns are appended to it, for the purpose of implementing multidimensional clustering.

The setup used in Oracle is extremely simple and is scalable to high chromatic numbers, i.e., many dimensions. It shows superior timing results when accessing table data that is already located in memory and bounds nicely as dimensionality increases.

\subsection{Inserts and Deletes}

Adding or deleting rows in the fact table cascades to the indexing structures, bitmap or clustering in the above. The responsibility to maintain the additional indexing structures therefore falls on the database management system in this case. This is more complicated when the digraph can also change, but the updating issues are similar to what is seen in data warehousing. If the digraph changes frequently then one should consider decoupling the clique structure and the fact table, in order to not cause to much processing during updates. There are many relational indexing techniques compatible with such a decoupling and used in data warehousing setups, we do not discuss these here.

\section{Additional Applications}
\label{sec:applications}

Now we demonstrate that the indexing is applicable to setups not just involving graph data. Both of the applications described may be used to resolve intersection and stabbing queries efficiently and are therefore interesting in their own right.

\subsection{Spatial Indexing}

Clique indexing techniques can be used to efficiently resolve stabbing and interval queries. Starting with a table containing or referencing intervals, one may define the set of data entries as the set of all endpoints of the intervals. The set-value function chosen for stabbing and interval queries may be selected as
$$
\begin{array}{rcl}
F(e*) &=& \{\mbox{Intervals}~I = (X,Y)~\mbox{in the table}\\
& &~~~\mbox{with}~Y > e* ~\mbox{and}~ X \leq  e*\}.
\end{array}$$
The purpose of this definition is not necessarily to materialize the set valued function. It may be regarded as a theoretical object, used only in the definition of an intersection graph that will then be efficiently colored using proper graph coloring techniques. This particular intersection graph, $\Int(F)$ can be properly colored extremely efficiently. Essentially, a simple geometric argument shows that the graph can be properly colored simply by ordering the nodes in an increasing order and then assigning the colors 1 through the chromatic number (or higher) sequentially and repeatedly to the ordered sequence of nodes. This requires an estimate of the chromatic number to be available to the system which is an additional step in the process. The dimensionality of this particular schema can be high, so it may have to be replaced with more than one schema, each referencing intervals of similar length, to keep the accumulated dimensionality low.

Once the intersection graph $\Int(F)$ has been properly colored and materialized in a clique indexing schema, e.g., a table called ``Clique\_F", an intersection query may be formed as follows. Assume that the ``node" column in ``Clique\_F" references closed interval objects with endpoints: node..x and node..y, and that $[a,b]$ is an arbitrary interval, also assume that $N*$ is the greatest data entry smaller than or equal to $b$ and, finally, that the data entry $N*$ has been assigned color ``n". In this case the intersection query: Select all intervals that intersect the closed interval $[a,b]$, is issued as:
\begin{itemize}
\item[-]
select node from Clique\_F where Cn = N*\\
union all\\
select node from Clique\_F where node..y between a and b
\end{itemize}
Each part of this query can be efficiently evaluated, for example, by using a bitmap index on the color column and a B+Tree index on the $node..y$ column.

\subsection{Homology Index}

Above, we created an indexing structure that uses endpoints of the intervals as data entries. This requires considerable processing when the interval data changes. Another approach is to use fixed reference intervals and introduce additional filtering to root out false positive hits. Consider the interval tree shown 
$$\begin{array}{ccccccccccccccccc}
\vline&\cdot&\cdot&\cdot&\cdot&\cdot&\cdot&\cdot&1&\cdot&\cdot&\cdot&\cdot&\cdot&\cdot&\cdot&\vline\\
\vline&\cdot&\cdot&\cdot&2&\cdot&\cdot&\cdot&\vline&\cdot&\cdot&\cdot&3&\cdot&\cdot&\cdot&\vline\\
\vline&\cdot&4&\cdot&\vline&\cdot&5&\cdot&\vline&\cdot&6&\cdot&\vline&\cdot&7&\cdot&\vline\\
\vline&~8~&\vline&~9~&\vline&10~&\vline&11~&\vline&12~&\vline&13~&\vline&14~&\vline&15~&\vline\\
\end{array}$$
This binary interval tree enumerates $2^n - 1$ intervals between $[0,1]$ where $n$ is the number of levels chosen, in the case shown we have $n=4$. Assuming we are interested in evaluating overlap queries, it is straightforward to pick the data entries as the intervals, numbered $1,2,3,\ldots,2^n - 1$ and set the set-valued function $F(i)$ as all intervals that overlap interval number $i$, e.g., $F(5) = \{1,2,5,10,11\}$. One quickly realizes that for this particular choice we have $$\chi(
\Int(F)) = 2^n - 1.$$ Since $1 \in F(i)$ always. This (usually) high chromatic number suggests that we should pick another function. The function, $G$, which we will use instead is defined as follows. The data entries used are the integer pairs $(p,q)*$ from the set:
$$\{(p,q) : q = 1,2,\ldots,n \mbox{~and~} p = 1,2,\ldots,2^q - 1 \}.$$
The set-valued function $G$, defined on this set and returning sets of intervals from the interval tree, is then determined by: $$\begin{array}{c}
k \in G(p,q)*\mbox{~if and only if~} \\ (k = p) \mbox{~or~} (2^q \leq 2p \leq k < 2^n \mbox{~and~} \Floor{\frac{k 2^q}{2^n}} = p).
\end{array}$$
Here $\Floor{x}$ is the floor function, i.e., returns the largest integer at most $x$. The intersection graph $\Int(G)$ can be properly colored with $n$ colors, since $G(p_1,q)\cap G(p_2,q) = \emptyset$ if $p_1 \neq p_2$. The coloring map is given by $$c(p,q)* = q\mbox{~and~}\chi(\Int(G)) = n.$$
This much more reasonable chromatic number ($n$ instead of $2^n - 1$) enables us to materialize the clique indexing schema into a compact relation, i.e., ``clique\_G". For example, if $n=20$ we obtain a table with 1,048,575 rows and 21 columns and containing no ``NULL" values.

The mechanism that assigns data entries to a given interval, in order to resolve an overlap query, in the interval graph is constructed as follows. Consider the $k$-th interval in the tree. It is located on level $L$, determined by $L = \Floor{\log_2(k)} + 1$. Recursively define a sequence $R_1,\ldots, R_L$ of length $L$ by $R_L = k,$ and then $R_{t-1} = \Floor{R_t / 2}$, note that $R_1 = 1$. Then the overlapping intervals are obtained as the disjoined union $G(R_L,L)*\cup G(R_{L-1},L)*\cdots\cup G(R_1,L)*$. So, referring back to the set $F(5) = \{1,2,5,10,11\}$, which we used to illustrate the earlier choice with n=4, we now obtain 
$$\begin{array}{rcl}
F(5) &=& G(5,3)*\cup G(2,3)*\cup G(1,3)*\\
 &=& \{5,10,11\}\cup\{2\}\cup\{1\}.
 \end{array}$$
In a database system this simply means that the intersection query: Find all intervals in the interval tree that intersect interval number $k$ in the tree, is issued as:
\begin{itemize}
\item[-]
select node from clique\_G where \\ cL in $(R_L,R_{L-1},\ldots,1).$
\end{itemize}
There is no need to store the data entries as pairs in the ``clique\_G" relation since the second value is just the color and therefore is the same within each column, i.e., it is enough to store the first number ``p" in the color columns, instead of $(p,q)*$. This explains why we write ``cL" in the above query for the L-th color column, instead of ``cL..p", to obtain the first value only.

Of course, this structure is also easily obtained by direct calculations. The advantage of materializing this compact schema, ``clique\_G", shown below for n = 4, is that it facilitates fast resolution of intersection queries when the intervals reference the interval tree.

$$\begin{array}{ccc}
\begin{tabular}{|c|c|c|c|c|}
  \hline
  Node & c1 & c2 & c3 & c4 \\ \hline
  1 & 1 & 1 & 1 &1 \\ \hline
  2 & 1 & 2 & 2 &2 \\ \hline
  3 & 1 & 3 & 3 &3 \\ \hline
  4 & 1 & 2 & 4 &4 \\ \hline
  5 & 1 & 2 & 5 &5 \\ \hline
  6 & 1 & 3 & 6 &6 \\ \hline
  7 & 1 & 3 & 7 &7 \\ \hline
\end{tabular} &
\begin{array}{l} \\ \& \\
\end{array} &
\begin{tabular}{|c|c|c|c|c|}
  \hline
  ~~~8~~~ & ~1~ & ~2~ & ~4~ & ~8~ \\ \hline
  9 & 1 & 2 & 4 &9 \\ \hline
  10& 1 & 2 & 5 &10\\ \hline
  11& 1 & 2 & 5 &11\\ \hline
  12& 1 & 3 & 6 &12\\ \hline
  13& 1 & 3 & 6 &13\\ \hline
  14& 1 & 3 & 7 &14\\ \hline
  15& 1 & 3 & 7 &15\\ \hline
\end{tabular}
\end{array}$$

\noindent
Our approach is to use the ``clique\_G" relation, in the Oracle 9.2i database, to build $n$ bitmap join indexes, one for each color to index a table that references features of the DNA. The indexing is then used to resolve queries involving the mapping of feature sets from one species to another.

\subsection*{Acknowledgments}

This research has been sponsored by deCODE Genetics, Inc. in Iceland.

\end{multicols}

\appendix
\section{Appendix: Testing}
\label{sec:appA}
\begin{figure}[t]
  $$\includegraphics[scale = 0.5]{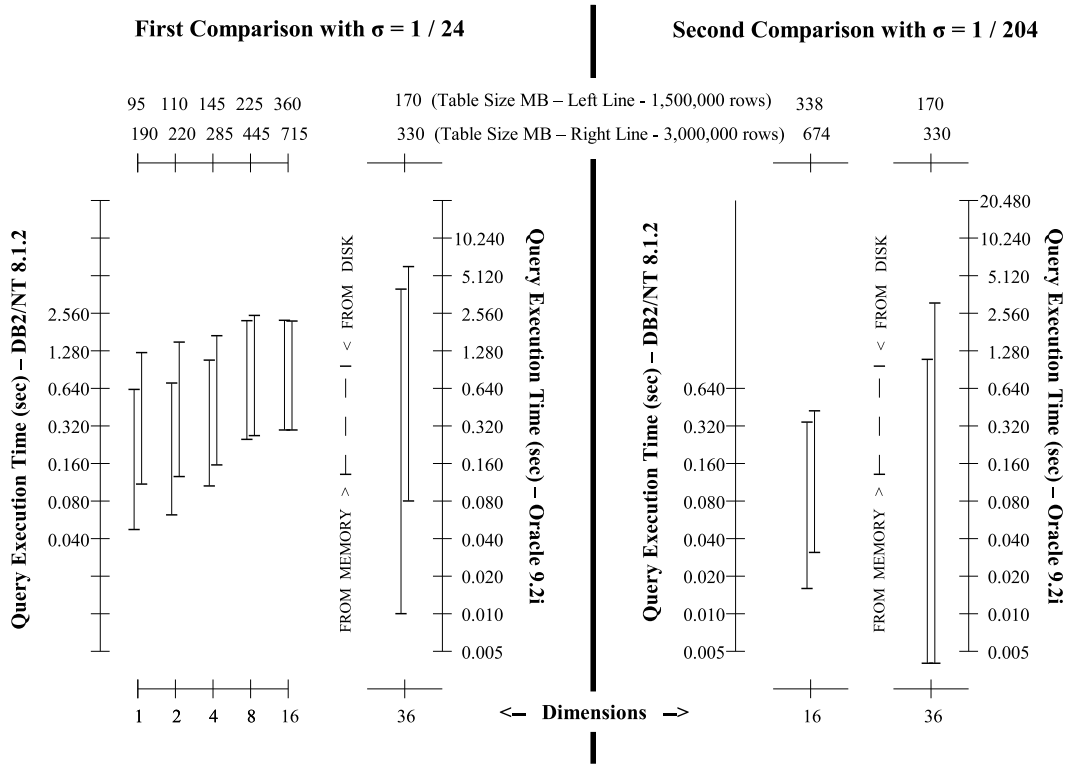}$$
  \caption{Comparisons with $\sigma$ equal to 1/24 and 1/204}\label{fig1}
\end{figure}
\begin{multicols}{2}
Here we document the testing described in section \ref{sec:comparisons}.

\subsection{Hardware and Data}
The computer used is a 2 processor Intel Pentium III 667MHz system each with integrated 256KB level 2 cache. It has a 512MB ECC RDRAM memory and all the database files are located on a WD 200GB, 7200 RPM disk with 8MB cache and 100MB/s ATA interface. The operating system is Windows 2000 Professional SP3. Care was taken to make sure that both databases, Oracle 9.2i and DB2 8.1.2 are optimally tuned and were able to take advantage of the limited resources available on this machine. The ``page size" was kept at 4KB in DB2. 

Two tables, one with approximately 1.5 million rows and the other with approximately 3 million rows reference the gene ontology, $\GOdag$, graph in a column called ``acc". The tables contain a number column also, called ``m", and other columns with various values, the referencing tables will be referred to as fact tables.

\subsection{The Gene Ontology Digraph}
The gene ontology acyclic digraph, $\GOdag$, is developed by the Gene Ontology Consortium, \cite{GO}. A greedy coloring algorithm may be used to properly color its intersection graph $\Int(D[\cdot])$ with 36 colors. The clique indexing schema $\clique(D[\cdot])$, denoted by clique\_GO in the queries below, therefore has 37 columns (Node, c1,c2,..., c36) and one row for each node in the digraph, i.e., over 11,000 nodes, following the convention established in section \ref{sec:generalized}.

\subsection{Fig. \ref{fig1}: Description and Queries}
Figure \ref{fig1} shows results from the first two comparisons provided and detailed below. In DB2's case we always use the color columns in the clique structure to cluster the fact tables using multidimensional clustering. In Oracle's case we always indexed the fact tables using bitmap join indexes derived from the color columns in the clique structure.

Referring back to Figure \ref{fig1}, the vertical axes show query execution time in seconds. The lower horizontal axes show the dimensionality of the system being tested. In DB2's case this means the number of color columns used to organize the fact tables using multidimensional clustering. In the first comparison shown, dimensionality equal to one indicates that the fact tables are clustered using color column ``c8'' only, dimensionality equal to 2 indicates that the clustering is organized by colors 7 and 8 and so on until the fact table is clustered simultaneously by all the color columns from 1 to 16 - which is the maximum supported by DB2. In the second comparison shown, color columns 9 through 24 were used as clustering columns simultaneously. In Oracle's case the dimensionality, i.e., 36, is simply the number of color columns in the clique relation. The Oracle setup is similar to a star-schema with 36 dimensions or dimension levels (independent), in which the clique indexing schema acts as a dimension table.

Each setup is executed both for a smaller fact table with approximately 1.5 million rows and a larger setup with approximately 3.0 million rows. The range of execution times for the smaller setup is always shown to the left and the range of execution times for the larger setup is always to the right. The time ranges are broad since they vary from the database having to read all the data from disk (cold database) to the execution times seen in repeated executions.

The top horizontal axes show the size of the tables in the database, both for the larger and the smaller fact table setup. As expected the size of the multidimensional clustered table increases substantially as it is organized by more and more dimensions. On the other hand, the star schema setup used in the Oracle database does not increase the size of the fact table as dimensionality increases, but the accumulated size of all the 36 bitmap join indexes, one for each color column, was only about 40MB in the Oracle 9.2i database.

The value of $\sigma$ shown for each comparison on figure \ref{fig1} is simply the ratio of rows in the fact table touched by the respective queries. So a value of $\sigma = 1/24$ means that about 5\% of the rows in the fact table satisfy the query, and the other value of $\sigma = 1/204$ indicates that only 0.5\% of the rows satisfy the second query. More precisely, figure \ref{fig1} shows execution times for the following queries:

\noindent First Comparison: For $\sigma = 1/24$, we select a node from the Gene Ontology graph assigned code `GO:0006810' representing `transport' (The directed movement of substances such as macromolecules, small molecules, ions into, out of, within or between cells). The query aggregates ``m" over all rows that reference this term or one of its many descendants in the digraph. As indicated about 5\% of the rows in the fact table satisfy this query. The node `GO:0006810' was assigned color 8 by the proper coloring algorithm applied to the intersection graph.
\begin{itemize}
\item[-] In DB2 the query used is of the form:
\\ $~$select sum(m) from MDC\_gotermfact 
\\ $~~~~~$where c8 = `GO:0006810'.\\ The necessary color columns included in the multidimensional clustering were added to the fact table.
\item [-] In Oracle the query was formed as:
\\ $~$select sum(f.m) 
\\ $~~~$from gotermfact f, clique\_GO d 
\\ $~~~~~$where d.node = f.acc and 
\\ $~~~~~~~~~~~~~~~~~$d.c8 = `GO:0006810'. \\
The ``clique\_GO" table is the clique indexing schema as explained above. Additionally, a hint (+index\_combine(f)) was inserted into the query to make sure that Oracle selected to use the available bitmap join indexes.
\end{itemize}

\noindent Second Comparison: For $\sigma$ = 1/204 we used the following nodes `GO: 0015171' (i.e., amino acid transporter activity), `GO:0015203' (i.e., poly-amine transporter activity), `GO:0015291' (i.e., porter activity) and aggregate over all rows in the fact table that reference a node which is the descendant of all these three nodes simultaneously. The nodes were assigned colors 13, 21 and 23 respectively by the coloring of the intersection graph.
\begin{itemize}
\item[-] This time, the query used for DB2 is as follows:
\\ $~$select sum(m) 
\\ $~~~$from MDC\_gotermfact\_9to24 
\\ $~~~~~$where c13 = `GO:0015171' and
\\ $~~~~~~~~~~~~~~~~~$c21 = `GO:0015203' and
\\ $~~~~~~~~~~~~~~~~~$c23 = `GO:0015291'.\\
The table is organized (clustered) by all the 16 color columns from c9 to c24.
\item[-] In Oracle the same query is formed as:
\\ $~$select sum(f.m) 
\\ $~~~$from gotermfact f, clique\_GO d 
\\ $~~~~~$where d.node = f.acc and 
\\ $~~~~~~~~~~~~~~~~~$d.c13 = `GO:0015171' and 
\\ $~~~~~~~~~~~~~~~~~$d.c21 = `GO:0015203' and 
\\ $~~~~~~~~~~~~~~~~~$d.c23 = `GO:0015291'.\\
Additionally, a hint was inserted to ensure the used of bitmap join indexes.
\end{itemize}

\end{multicols}

\today

\begin{thebibliography}{10}

\bibitem{GeirAgust1}
\newblock Geir Agnarsson and \'{A}g\'{u}st Egilsson.
\newblock {\em On vertex coloring simple genetic digraphs.}
\newblock To appear (Congressus Numerantium, A conference journal on numerical themes).

\bibitem{GeirAgust2}
\newblock Geir Agnarsson, \'{A}g\'{u}st Egilsson and Magn\'{u}s M.~Halld\'{o}rsson.
\newblock {\em Proper down-coloring simple acyclic digraphs.}
\newblock To appear (Springer Lecture Notes in Computer Science).

\bibitem{Variant}
\newblock Patrick O'Neil and Dallan Quass.
\newblock {\em Improved Query Performance with Variant Indexes.}
\newblock In SIGMOD 1997, Proceedings ACM SIGMOD International Conference on Management of Data, 1997.

\bibitem{Sarawagi}
\newblock Sunita Sarawagi and Michael Stonebraker.
\newblock {\em Efficient Organization of Large Multidimensional Arrays.}
\newblock 10th International Conference on Data Engineering, IEEE 1994.

\bibitem{Zhao}
\newblock Yihong Zhao, Prasad M. Deshpande and Jeffrey F. Naughton.
\newblock {\em An Array-Based Algorithm for Simultaneous Multidimensional Aggregates.}
\newblock In SIGMOD 1997, Proceedings ACM SIGMOD International Conference on Management of Data, 1997.

\bibitem{Framework}
\newblock Marcel Kornacker, Mehul Shah and Joseph Hellerstein.
\newblock {\em An Analysis Framework for Access Methods.}
\newblock Technical Report ``CSD-99-1051" available from http://db.cs.berkeley.edu/papers/, June 1999, UC Berkeley.

\bibitem{Gisli}
\newblock Brian F. Cooper, Neal Sample, Michael J. Franklin, Gísli R. Hjaltason and Moshe Shadmon.
\newblock {\em A Fast Index for Semistructured Data.}
\newblock In proceedings of the 27th International Conference on Very Large Databases (VLDB 2001), Roma, Italy, September 2001, 341–350.

\bibitem{RI}
\newblock Hans-Peter Kriegel, Marco P\"otke and Thomas Seidl.
\newblock {\em Object-Relational Indexing for General Interval Relationships.}
\newblock Proc. 7th Int'l Symposium on Spatial and Temporal Databases (SSTD'01), Lecture Notes in Computer Science, Springer Verlag, 2001.


\bibitem{GO}
\newblock The Gene Ontology Consortium.
\newblock {\em Gene Ontology: tool for the unification of biology.
The Gene Ontology Consortium (2000).}
\newblock Nature Genet, {\bf 25}, 25 -- 29, (2000).

\bibitem{Ab-Bu-Su}
\newblock Serge Abitebboul, Peter Buneman and Dan Suciu.
\newblock {\em Data on the Web, From Relations to Semistructured Data and XML.}
\newblock Morgan Kaufmann Publishers (2000).

\bibitem{Oracle-Star}
\newblock An Oracle White Paper.
\newblock {\em Key Data Warehousing Features in Oracle9i:
A Comparative Performance Analysis.}
\newblock Available online from Oracle at
{\tt http://otn.oracle.com/products/oracle9i/pdf/o9i\_dwfc.pdf},
(September 2001).

\bibitem{IBM-Cluster}
\newblock B.~Bhattacharjee, L.~Cranston, T.~Malkemus, and S.~Padmanabhan.
\newblock {\em Boosting Query Performance: Multidimensional Clustering}
\newblock DB2 Magazine, Quarter 2, 2003, Vol. 8, Issue 2. Also, available online at
 {\tt http://www.db2mag.com}

\bibitem{DWest}
\newblock Douglas B.~West.
\newblock {\em Introduction to Graph Theory.}
\newblock Prentice Hall, second edition, (2001).

\end{thebibliography}
\end{document}